\documentclass[
  reprint,            
  superscriptaddress, 
  amsmath,amssymb,    
  aps,                
  floatfix,           
]{revtex4-2}          

\usepackage{graphicx}
\usepackage{physics}
\usepackage{hyperref}
\usepackage{xcolor}
\usepackage{url}
\usepackage{booktabs}
\usepackage{tikz}
\usetikzlibrary{shapes.geometric}

\definecolor{bluel}{HTML}{B8D8EC}
\definecolor{bluem}{HTML}{7FA8CC}
\definecolor{blued}{HTML}{3E6E9E}
\definecolor{bluex}{HTML}{12355E}

\newcommand{\key}[4][4.4pt]{%
  \tikz[baseline=-0.55ex]{%
    \draw[#2, line width=#4] (0,0) -- (1.0em,0);
    \node[#3, draw=#2, fill=white, line width=#4,
          inner sep=0pt, minimum size=#1] at (1.4em,0) {};
    \draw[#2, line width=#4] (1.8em,0) -- (2.8em,0);}}

\begin{document}

\newcommand{\rlh}{\rightleftharpoons}









\title{Local energetic coupling enhances the expressivity of chemical computation}




\providecommand{\CSL}{Complex Systems Lab, Universitat Pompeu Fabra, 08003, Barcelona, Spain.}
\providecommand{\Dayhoff}{Dayhoff Labs, Inc., Cambridge, MA 02138}

\author{Marco Tuccio}
\email{marco@dayhofflabs.com}
\affiliation{\Dayhoff}
\affiliation{\CSL}

\author{Jason W. Rocks}
\email{jason@dayhofflabs.com}
\author{Joshua E. Goldford}
\email{josh@dayhofflabs.com}
\affiliation{\Dayhoff}

\begin{abstract}
Living systems compute with chemistry by mapping environmental signals onto specific internal chemical states. Despite recent advances in molecular programming, it remains unclear which physicochemical features control the computational expressivity of chemical systems. Here we inverse-design thermodynamically consistent chemical reaction networks whose steady-state response to an environmental input computes a target nonlinear function. Using implicit differentiation we train the free-energy landscape directly: standard chemical potentials, transition-state energies and thermodynamic drives. Increasingly large networks generated by elementary ligation and cleavage steps fit increasingly complex nonmonotonic polynomial functions, with expressivity scaling logarithmically with network size, predicted primarily by the number of reactions. Training individual energetic parameter classes reveals that internal thermodynamic drives, capable of breaking detailed balance, dominate trainability, with comparable performances achieved only by pairs of parameter classes. These results identify nonequilibrium drive as the most effective single resource for steady-state computational expressivity in chemical reaction networks.
\end{abstract}

\maketitle

\begin{figure*}[t]
    \centering
    \includegraphics[width=\linewidth]{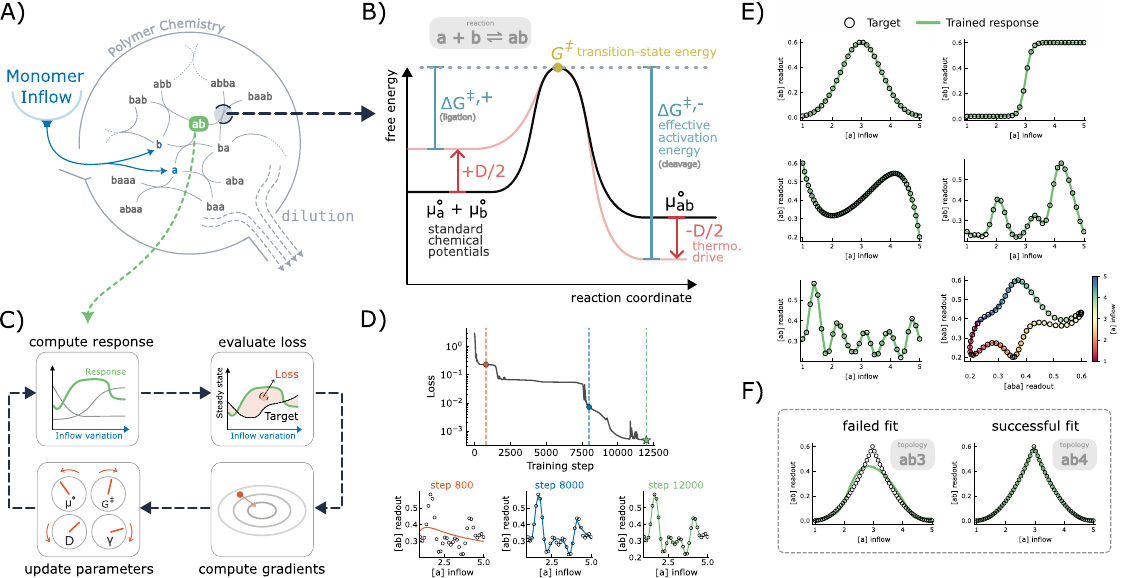}
    \caption{\textbf{(A)} Cartoon of a ligation--cleavage polymerization network in a
    Continuously Stirred Tank Reactor (CSTR). The network contains every sequence
    that can be assembled from the specified alphabet size $A$ up to a maximum length $L$, together with all ligation and cleavage reactions among them.
    \textbf{(B)} Parameterization of the reaction kinetics in terms of
    standard chemical potentials $\mu_i^\circ$, transition-state energies
    $G^\ddagger_\alpha$, and thermodynamic drives $D_\alpha$, which together set
    the effective activation barriers $\Delta G^{\ddagger,\pm}_\alpha$ in the
    ligation ($+$) and cleavage ($-$) directions.
    \textbf{(C)} Schematic of the training loop. The steady-state response is
    computed as a function of the monomer reservoir concentration, and the mean
    squared error against the target function is backpropagated to the reaction
    parameters, which are updated with Adam.
    \textbf{(D)} A typical training trajectory for an $abcdef3$ network with $N=258$ species and $R=462$ reactions, fit to a highly oscillatory target: the loss over training steps, with snapshots of the response at selected steps.
    \textbf{(E)} Example topology/target pairs, including several smooth nonmonotonic targets and a parametric fit with one input and two outputs tracing the von Neumann elephant.
    \textbf{(F)} A quadratic tent that \textit{ab3} fails to capture is fit almost perfectly by \textit{ab4}.}
    \label{fig:fig1}
\end{figure*}

Cells map the molecular composition of their environment onto specific internal chemical states, a form of computation carried out by chemistry itself. Bacteria track chemoattractant gradients and adapt their sensitivity through receptor methylation \cite{Berg1972-mk, Barkai1997-er}, quorum-sensing circuits convert cell density into coordinated gene expression \cite{Waters2005-gd}, and morphogen gradients are read out into distinct cell fates during development \cite{Wolpert1969-sn, Gregor2007-xq}. The same behavior can be built from scratch in synthetic systems, including engineered gene circuits such as toggle switches and oscillators \cite{Gardner2000-yb, Elowitz2000-tf}, DNA strand-displacement networks that implement logic and even neural-network classification \cite{Qian2011-ot, Qian2011-qy, Cherry2018-ly, Cherry2025-ml}.

Across both natural and engineered examples, the same questions arise. What are the physicochemical rules and constraints that govern such chemical computation? Which input-output behaviors are or are not computable by a given chemical system, and what determines the limits on expressivity? Answering these questions would not only enable the design of more capable synthetic circuits, but also reveal the physicochemical constraints that have guided the evolution of computation in living systems.

To answer these questions faithfully, two requirements must be met: a model must be grounded in the real physical chemistry of reactions, and it must admit a characterization of what the system can and cannot compute. Existing work tends to secure one at the cost of the other. Stochastic thermodynamics derives physically grounded bounds on the energetic cost of chemical computation~\cite{Wolpert2024-iw, Wolpert2025-gh, Liang2026-br}, but says nothing about the expressiveness of that computation. For Markov jump processes, which approximate linear unimolecular chemical reaction networks, expressivity scales with nonequilibrium driving
and is set by the number of reactions a single input
modulates~\cite{Floyd2025-iv}. Chemical Reaction Network Theory studies steady-state structure directly and could in principle predict which responses a network admits~\cite{Feinberg2019-hj, Horn1972-du}, but its closed-form results hold only for restricted structural classes~\cite{Feinberg1987-dg}. The very richness that makes a chemistry realistic is what puts it beyond the reach of theorems and closed-form results, calling for a computational and numerical approach instead. The two dominant computational approaches make opposite compromises: constraint-based methods such as flux balance analysis \cite{Orth2010-qt} reach genome scale by discarding kinetics altogether, while kinetic models \cite{Srinivasan2015-gs} retain rate laws at the cost of parameters that are difficult to identify and rarely tractable beyond small pathways.

Molecular programming shows what idealized chemistries can be built to do~\cite{Soloveichik2008-as, Soloveichik2010-cw, Qian2011-qy, Fages2017-cp, Cardelli2018-up, Anderson2021-yl, Cherry2025-ml, Poole2025-yx}, but is often detached from real physical chemistry in four main ways. First, models are often rate-independent, with behavior determined by network topology alone~\cite{Vasic2022-sx, Soloveichik2010-cw}; where kinetics are considered, networks are assembled by hand from computational primitives in the manner of electronic circuits. Second, reactions carry no notion of transition-state energies, chemical potential differences or free-energy landscapes~\cite{Vasic2020-vn, Qian2011-qy, Salehi2017-zn}, and are most often arrays of irreversible and autocatalytic steps~\cite{Poole2025-yx, Soloveichik2008-as, Cardelli2018-up}. Third, such reactions are not thermodynamically consistent, and work is implicitly injected without bound~\cite{Lee2025-dd, Cardelli2018-up, Mordvintsev2023-iv}. Fourth, inputs are typically encoded as initial concentrations that set the starting point of the ODE dynamics, without sustained coupling to the environment~\cite{Qian2011-qy, Salehi2017-zn, Fages2017-cp, Vasic2020-vn, Cherry2018-ly, Mordvintsev2023-iv}. Individually these are convenient modeling choices, but together they place the model outside the physics whose limits we set out to measure. A network that injects work without bound or ignores detailed balance cannot speak to what physical chemistry permits, nor can it be mapped onto a reaction system that could be realized in a lab. Restoring these constraints is what makes both the theoretical question and an experimental test well-posed.

Here we remove these idealizations and restore the relevant physical constraints, developing a physicochemically interpretable framework to probe the limits of computation in CRNs. To achieve a desired computational behavior, we train chemical reaction networks (CRNs) to exhibit specific chemical responses to environmental concentrations. We fit the kinetic rate constants indirectly, through a decomposition into physically interpretable energetic quantities: transition-state energies, standard chemical potentials, and thermodynamic biases. Fitting in this representation keeps every learned network thermodynamically consistent while making explicit which physical degrees of freedom the computation actually uses. We find that the capacity to fit harder targets follows an approximately log-linear relationship with network size. Studying the effects of individual physicochemical parameters highlights that thermodynamic free energy driving is the dominant source of this expressivity.

\section{Setup}

We interpret CRN computing as realizing some steady-state concentration profile against changing environmental concentrations. Formally, we ask a CRN to realize a target function $y=f(x)$, where $x$ is an environmental concentration we control freely and $y$ is the steady-state concentration of a chosen species. Training consists of tuning the reaction network until its response matches $f$.

In order to keep the topological features of our reaction networks consistent while scaling their size, we focus on polymerization networks~\cite{Moyer2021-se}, a physically grounded abstraction of the reversible condensation--hydrolysis chemistry underlying biological heteropolymers such as peptides and oligonucleotides, in which a bond-forming condensation is opposed by its hydrolytic reverse~\cite{Higgs2016-ae}. A polymerization network $abc\dots L$ is defined by an alphabet of monomers $\{a, b, c, \ldots\}$ of size $A$ together with a maximum polymer length $L$, and it comprises every ligation and cleavage reaction among the sequences that can be assembled from that alphabet up to $L$ [Fig.~\ref{fig:fig1}(A)]. A ligation joins two sequences end to end into a longer one, and the reverse cleavage splits a sequence back into two contiguous fragments. The network contains one reversible reaction for every such split. The smallest nontrivial network, $ab2$, contains the monomers $a$ and $b$ along with $aa$, $ab$, $ba$, and $bb$. Because concatenation is order dependent, a single reactant pair can yield distinct products. The pair $a+b$ produces both $ab$ and $ba$, in ratios set by the kinetics of {$a+b\rlh ab$} and {$a+b\rlh ba$}. Conversely, a given polymer is reached by every ligation of shorter fragments that reproduces its sequence, and all such paths are present in the topology. Network size grows in alphabet size and maximum length, so networks spanning several orders of magnitude in size follow from the same generative primitives and share the same fundamental topological features.

We describe CRNs consisting of a collection of $N$ chemical species (index $i$) interacting via a set of $R$ reversible reactions (index $\alpha$) as a system of ODEs,
\begin{equation}
    \dv{c_i}{t} =  \sum_\alpha S_{i\alpha}J_\alpha+ J_i^{\mathrm{ext}}
\end{equation}
where $c_i$ is the concentration of species $i$, $J_\alpha=J^+_\alpha-J^-_\alpha$ is net flux of reaction $\alpha$ with forward/backward fluxes $J^+_\alpha$/$J^-_\alpha$, and $S_{i\alpha}$ is the stoichiometric matrix. 

To capture continuous environmental coupling of inputs we employ a Continuously Stirred Tank Reactor (CSTR) setting. Our system is supplied with monomers via inflow, while all species are diluted at the same rate, conserving overall volume. We model both with the external flux
\begin{equation}
    J_i^{\mathrm{ext}} = \gamma \qty(c^{\mathrm{ext}}_i - c_i)
\end{equation}
where $\gamma$ is the flow (and dilution) rate per unit volume into and out of the system, and $c^{\mathrm{ext}}_i$ is the concentration of each species in the inflow source. Monomers are supplied by the external source simply by setting their $c^{\mathrm{ext}}_i > 0$. 

All reactions in our setup obey mass-action kinetics,
\begin{equation}
    J_\alpha^\pm = k^\pm_\alpha \prod_{i=1}^N c_i^{\nu^\pm_{i\alpha}},
\end{equation}
where $k_{\alpha}^{\pm}$ are the forward and backward rate constants and $\nu_{i\alpha}^\pm \geq 0$ are the stoichiometric coefficients of species $i$ among the reactants of the $\pm$-direction of reaction $\alpha$; thus $\nu^+$ collect the left-hand side and $\nu^-$ the right-hand side, and $S_{i\alpha} = \nu^-_{i\alpha}-\nu^+_{i\alpha}$ is product minus reactants. Throughout, we orient every reaction so that the forward ($+$) direction is ligation and the backward ($-$) is cleavage.

Directly training kinetic rates $k^\pm_\alpha$ would let us tune specific thermodynamic features of the system without being able to untangle their exact energetic source and our goal is to investigate how these different sources contribute to training. To do this, we express kinetic rates in terms of effective activation free energies \cite{Hanggi1990-ei}
\begin{equation}
    k^\pm_\alpha = k_0 \exp \qty(-\Delta G^{\ddagger,\pm}_\alpha \over  k_B T)
\end{equation}
where $k_0 = c_0/\tau$ where $\tau$ is a standard timescale we set to $1$. The effective energy barriers $\Delta G^{\ddagger,\pm}_\alpha$ decompose as follows \cite{Hill2012-ed, Schmiedl2006-iv, Liebermeister2010-nz}:
\begin{equation}
    \Delta G^{\ddagger,\pm}_\alpha = G^\ddagger_\alpha - \sum_{i=1}^N \nu_{i\alpha}^\pm\, \mu_i^\circ \mp \frac{1}{2}D_\alpha.
   \label{eq:kineticsparam}
\end{equation}
In this decomposition, $\mu_i^\circ$ are the standard chemical potentials, or free energies of formation for each species involved in the reaction. The terms $G^\ddagger_\alpha$ and $\mp \frac{1}{2}D_\alpha$ are reaction-specific catalytic parameters. The first parameter $G^\ddagger_\alpha$ is the transition-state energy, or standard free energy of the activated complex, which sets absolute free energy at the top of the reaction barrier, effectively controlling the overall intrinsic speed of the reaction. While this term does not contribute to the reaction steady-state in an equilibrium closed system, it does contribute in a nonequilibrium open CSTR setup where dilution imposes a competing timescale \cite{Rao2016-bc}, so the steady state depends on each reaction's speed relative to $\gamma$. The second parameter $D_\alpha$ is the nonconservative thermodynamic drive or bias, representing energetic coupling (e.g., ATP hydrolysis), biasing the reaction in one direction or another. Together, $\mu_i^\circ$, $G^\ddagger_\alpha$, and $D_\alpha$, provide an explicit interpretation of the rate constants in terms of each reaction's energetic landscape [see Fig.~\ref{fig:fig1}(B)].

We are interested in finding parameter regions where the steady-state response of the network matches a target function $f(x)$. The procedure to find them is depicted in Fig.~\ref{fig:fig1}(C) and amounts to training a CRN. We evaluate its output and update its parameters to reduce the loss given by the error between the evaluated output and the desired target. CRNs typically settle onto a fixed point at which every concentration $c_i^*$ is constant in time. In our framework, the input variable is the reservoir concentration $x = c^{\mathrm{ext}}_{\mathrm{in}}$ of a designated monomer, with all other monomer reservoir concentrations held fixed. Our output is the steady-state concentration $y^* = c^*_{\mathrm{out}}$ of a designated polymer produced by the system. Our fit parameters consist of all energetic terms and the dilution rate, $\theta = (\{G^{\ddagger}_\alpha\}, \{\mu^\circ_i\}, \{D_\alpha\}, \gamma)$.

We sweep the input over a grid of $K$ values $\{x_k\}$ and compare the resulting response to the target at each of them. The mean squared error defines the loss,
\begin{equation}
    \mathcal{L}_{\mathrm{fit}}(\theta) = \frac{1}{K}\sum_k \qty[ y^*(x_k;\theta) - f(x_k)]^2,
\end{equation}
which we minimize with a standard Adam optimizer, updating $\theta$ at each training step. Two choices make the required gradients $\partial\mathcal{L}/\partial\theta$ cheap to obtain. First, we evaluate the steady state by root-finding rather than by integrating the often stiff CRN dynamics. Second, rather than backpropagating through the root-finder, we use the implicit function theorem to derive exact gradients~\cite{Blondel2021-wr}. We further regularize the loss to bias parameters toward regions where the fixed point is dynamically stable. Details can be found in the Appendix~\ref{apx:training}.

Figure~\ref{fig:fig1}(D) shows a typical training progression, and Fig.~\ref{fig:fig1}(E) demonstrates the method can train CRNs to reproduce a wide variety of functional forms, including Gaussian bumps (band-pass filters), step functions (hypersensitivity), and more complicated nonmonotonic targets. A single input read out along two channels traces out a von Neumann elephant parametric curve~\cite{Mayer2010-if} where readout species $aba$ and $bab$ are trained on the single input $ab$.

\begin{figure}[t]
    \centering
    \includegraphics[width=\linewidth]{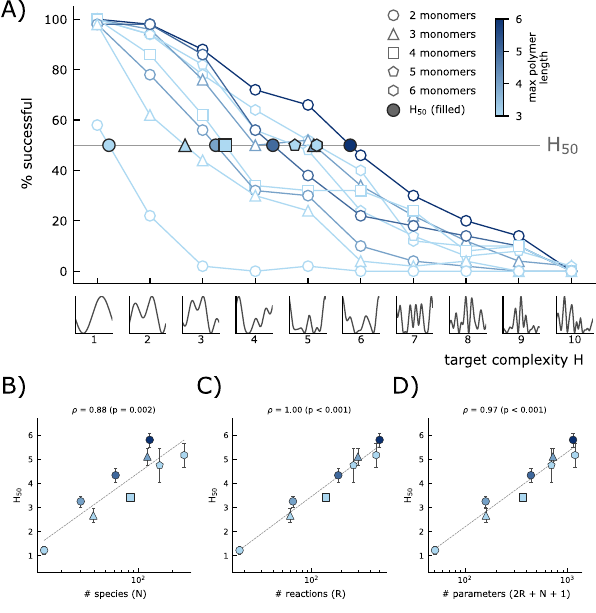}
    \caption{
\textbf{(A)} Success rate against target complexity $H$ for each of the nine polymerization networks containing up to ${\sim}500$ reactions. Targets are randomly generated squared trigonometric polynomials built from harmonics up to $H$ with random magnitudes and phases and correspond to target complexity; representative targets are drawn below the axis. In every run the input is the reservoir concentration of monomer $a$ and the output is the steady-state concentration of $ab$. Each point aggregates independent experiments for each topology against a set of $50$ random targets generated for each harmonic (shared across all topology for each target complexity) resulting in $4500$ training runs in total. A run counts as a success when it reaches $R^{2}\geq0.98$ and survives a quasistatic verification of the dynamics (Appendix~\ref{apx:odecheck}). Marker shape encodes alphabet size $A$ and color encodes maximum polymer length $L$, so the two generative axes we scale can be read independently. Filled markers give $H_{50}$, the complexity at which the success rate crosses $50\%$, interpolated between adjacent integer values of $H$.
\textbf{(B--D)} $H_{50}$ against three measures of network size: the number of species $N$ \textbf{(B)}, the number of reactions $R$ \textbf{(C)}, and the number of trainable parameters $2R+N+1$ \textbf{(D)}, the latter comprising one $G^\ddagger_\alpha$ and one $D_\alpha$ per reaction, one $\mu^\circ_i$ per species, plus the flow rate $\gamma$. Capacity tracks size in all three cases, most tightly for the reaction count (Spearman $\rho=1.00$), then the parameter count ($\rho=0.97$), then the species count ($\rho=0.88$). Dashed lines are log-linear fits; vertical error bars propagate the binomial uncertainty in the success rates into $H_{50}$.
\vspace{-30pt}
}
    \label{fig:fig2}
\end{figure}

\section{Expressivity versus network size}
\label{sec:scaling}

Not every network fits every target function. For example, Fig.~\ref{fig:fig1}(F) shows $ab3$ failing to fit the quadratic tent that is almost perfectly fit by $ab4$. The specific functions a given CRN can realize depend on the details of a network's topology and its compatibility with a desired target function.

We probe this dependence systematically by matching increasingly large reaction networks against increasingly complex targets. We vary alphabet size $A$ and maximum polymer length $L$ of networks, obtaining $9$ topologies of increasing size but under $\sim500$ reactions, listed in Table.~\ref{tab:architectures}. Every polymerization network contains the reaction $a+b \rlh ab$, so throughout we fix $a$ as the input monomer and $ab$ as the output polymer, isolating the effect of topology from the choice of input--output pair (Appendix~\ref{apx:readoutloss}). We then generate a set of randomly generated squared trigonometric polynomials, composed of $10$ classes of difficulty. Each target is built by summing harmonics terms up to a highest harmonic $H$, each with randomly drawn magnitudes and phases, then squaring the total. Squaring keeps the target positive and therefore admissible as a concentration readout. Target complexity is thus controlled by $H$, where lower $H$ gives simpler targets and higher $H$ superimposes more modes into more complicated targets (Appendix~\ref{apx:targets}). Examples appear along the $x$-axis of Fig.~\ref{fig:fig2}(A).

For each target complexity class $H$ we generate $50$ targets and run all topologies against the same targets for each class, for a total of $4500$ training runs. A run counts as a success when it reaches a coefficient of determination $R^2 \geq 0.98$ and survives a quasistatic verification of the dynamics (Appendix~\ref{apx:odecheck}). This verification is necessary because the root finder locates steady states independently at each input value and is blind to their stability and connectivity. A fitted response can therefore rest on unstable points rather than attractors, or stitch together coexisting branches of a multistable system that the true dynamics would not necessarily traverse. To exclude both failure modes we integrate the full dynamics along a quasistatic sweep of the input and retain only responses lying on a single, continuously connected branch. For each target complexity class $H$ and each topology, experiments yield a success rate: the fraction of targets at complexity $H$ that a topology fits [Fig.~\ref{fig:fig2}(A)]. We measure the computational expressivity of a topology as $H_{50}$: the target complexity class at which success crosses the $50\%$ probability, a continuous value found via interpolation (Appendix~\ref{apx:capacityestimation}).

Formulated this way, we find that expressivity tracks network size almost perfectly. The correlation is strongest with the number of reactions $R$, where the ranking is exact (Spearman $\rho = 1.00$), and nearly as strong with the number of trainable parameters $2R+N+1$ ($\rho = 0.97$), while the number of chemical species $N$ correlates somewhat less tightly ($\rho = 0.88$)~[Fig.~\ref{fig:fig2}(B,C,D)].

\section{Freezing parameter classes}

\begin{figure}[t]
    \centering
    \includegraphics[width=\linewidth]{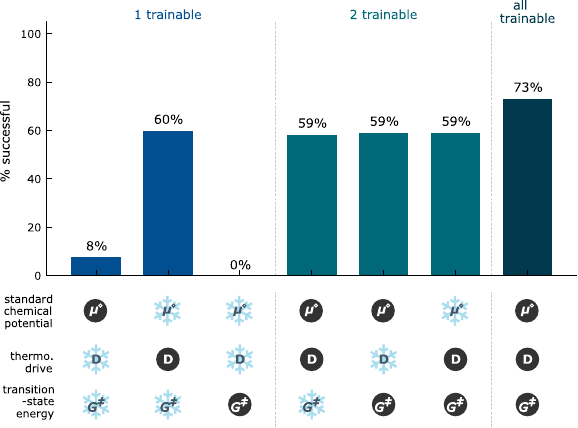}
    \caption{
Parameter-freezing study on a fixed $abc3$ topology with $N = 39$ species and $R = 60$ reactions, against an ensemble of $128$ random targets at target function complexity $H=2$. Each bin corresponds to one trainable subset of the three parameter classes: frozen classes (snowflake) are set to zero, while the remaining classes (black) are randomly initialized near zero and trained; the
flow rate $\gamma = 1$ is frozen throughout. Bars give the success rate, defined as in Fig.~\ref{fig:fig2} ($R^{2}\geq0.98$ together with a quasistatic verification of the dynamics). Among the single-class bins only the thermodynamic drives $D_\alpha$ train well ($60\%$), against $8\%$ for the standard chemical potentials $\mu^\circ_i$ and no successes for the transition-state energies $G^\ddagger_\alpha$; $D_\alpha$ is the only class that is both per-reaction and antisymmetric, and so the only one that can break detailed balance on an individual reaction. Adding a second class to $D_\alpha$ brings no further benefit, but the $(\mu^\circ_i, G^\ddagger_\alpha)$ bin,
with $D_\alpha$ frozen, matches the drive-only bins at ${\sim}60\%$. The fully trainable control performs best at $73\%$.}
    \label{fig:fig3}
\end{figure}

The tight correlation between expressivity and number of trainable parameters raises the question of which parameters carry it. The three classes are not interchangeable: they enter the kinetics differently (eq.~\ref{eq:kineticsparam}) and are constrained in different ways. The standard chemical potentials $\mu^\circ$ are assigned per species rather than per reaction, so a single $\mu_i^\circ$ is shared across every reaction that species participates in and cannot be tuned for one reaction in isolation. The transition-state energies $G^\ddagger_\alpha$ are per-reaction but enter both directions equally, setting reaction speed rather than thermodynamic bias. Only the drive $D_\alpha$ is both per-reaction and antisymmetric, making it the only class that can break detailed balance on an individual reaction.

We test this with freezing studies. We train a fixed topology $abc3$ against an ensemble of 128 random targets at $H=2$ harmonics. For each trainable subset of the three classes, the frozen classes are initialized to zero while the remaining classes are randomly initialized and trained, with a fully trainable control [Fig.~\ref{fig:fig3}]. For each bin we measure trainability as the percentage of successful training rounds among the total, with success always referring to passing ODE quasistatic verification and having a sufficiently good fit with $R^2 \geq 0.98$. We find that thermodynamic drives alone $D_\alpha$ carry almost all of the trainability, reaching a $60\%$ success rate compared to $8\%$ for the standard chemical potentials $\mu^\circ_i$ and no successes for the transition-state energies $G^\ddagger_\alpha$. For the pairwise studies, where we unfreeze two classes, both experiments where $D_\alpha$ is unfrozen exhibit no benefit from the addition of a second parameter class. However, the case where $D_\alpha$ is the only frozen parameter class and $\mu^\circ_i$ and $G^\ddagger_\alpha$ are both trainable, we achieve the same $59\%$ success rate as the studies including $D_\alpha$. The control where all three parameter classes are trainable performs best at $73\%$.

\section{Discussion}

We have shown that physically-parameterized, fixed-topology polymerization networks, with no catalysis or autocatalysis, can be tuned to reproduce a wide range of smooth steady-state responses by adjusting reaction energies alone. We do this via inverse-design of CRNs based on implicit differentiation through steady state solutions of a root-finding algorithm, which avoids backpropagating through stiff ODE integration \cite{Mordvintsev2023-iv, Lee2025-dd}. We show fitting of various one-dimensional maps of increasing difficulty and also train a network against the von Neumann elephant~\cite{Mayer2010-if} suggesting that our framework extends past scalar input-output maps to vector-valued target responses, which is where objectives like homeostasis and balanced growth lie.  
We show that the capacity to learn more complex responses scales logarithmically well with network size. However, we cannot identify more particular structural features as responsible for expressivity, since specific topological attributes are collinear with increasing network size. Future work could explore different reaction network generating primitives that enable decoupling of topological attributes of interest. 

The parameter-freezing studies in Fig.~\ref{fig:fig3} show that what enables a network to compute is the capacity to drive its reactions out of equilibrium and not the fine-tuning of individual transition state energies. If this holds more generally, acquiring a source of thermodynamic drive---such as a coupled energy carrier---would matter more for the emergence of chemical computation than precise control over reaction barriers. We should note that this result may primarily reflect our training target being a steady-state concentration; transient objectives may instead induce stronger kinetic than thermodynamic dependence when fitting specific time-dependent functional responses.

Future translational work may use the empirical findings here as design principles for engineering CRNs for bio-sensing and computational tasks. Combining CRNs constructed from more realistic chemical reaction mechanisms with recently developed machine learning tools for enzyme discovery \cite{Rocks2026-md} and design \cite{Ahern2025-ag} may provide a suitable design suite to guide experiments. Interestingly, experimental systems consistent with the chemistry model presented here---with precise tuning of dissipation and barrier heights---may be achievable with amino acid ligation enzymes (E.C.\ 6.3.2.-), which conjugate amino acids (or short peptides) with single amino acids in a reaction coupled to ATP hydrolysis. Leveraging the kinetic diversity of orthologs within this enzyme family across different substrates may enable the construction of real-world, driven chemistries with tunable properties.

Several extensions follow naturally. Template-directed chemistries would extend the framework to origins-relevant networks that couple informational polymers to their own synthesis \cite{Adamski2020-ci}, while adding explicit catalytic and autocatalytic edges would connect it to autocatalytic-set theory and its reflexive, self-maintaining structures (RAFs) \cite{Steel2019-si}, letting us ask which physical parameters govern autocatalytic gain. Objectives such as homeostasis, balanced growth, and error correction would test whether the same thermodynamic levers shape function beyond input--output fitting \cite{Goldford2018-nq, Murugan2012-mx, Ravasio2026-aw}. More broadly, future work can address whether more complex structural, topological or physico-chemical features of a network, such as compartmentalization, enable more expressive computation. Although our study establishes the existence of expressive solutions, it remains unclear whether those found by gradient descent are also accessible under evolutionary dynamics.

\begin{acknowledgments}
We thank Pankaj Mehta and Artemy Kolchinsky for valuable discussions. We thank the Dayhoff Labs team for the continuous feedback throughout this work. 
\end{acknowledgments}

\bibliography{library}

\appendix

\section{Chemical Reaction Network Modeling}
\label{apx:crn}



Below, we show the detailed mathematical construction of our systems with Table.~\ref{tab:symbols} summarizing all the notation used.

\subsection{Mass-Action Dynamics in a Continuously Stirred Tank Reactor}
\label{apx:dynamics}

Consider a network containing \(N\) chemical species and \(R\) reversible reactions. Species are indexed by \(i=1,\ldots,N\), and reactions are indexed by \(\alpha=1,\ldots,R\). Concentrations are measured relative to a standard reference concentration \(c^\circ\), energies relative to the thermal energy \(k_BT\), and times relative to $\tau = h/k_BT$ where \(h\) is Planck's constant and \(T\) is the reactor temperature (kept fixed across all simulations).

\begin{table}[t]
    \caption{
    Symbols used in the CRN formulation. Here \(\mathsf{C}\),
    \(\mathsf{E}\), and \(\mathsf{T}\) denote concentration, energy, and time. 
    Quantities marked \(1\) are dimensionless.
    }
    \label{tab:symbols}
    \footnotesize
    \begin{ruledtabular}
    \begin{tabular}{lll}
        Symbol & Meaning & Dimension \\
        \hline
        \(A\) & Alphabet size & \(1\) \\
        \(L\) & Maximum polymer length & \(1\) \\
        \(R\) & number of reversible reactions & \(1\) \\
        \(N\) & number of species & \(1\) \\
        \(K\) & number of input grid points & \(1\) \\
        \(\nu^\pm_{i\alpha}\) & stoichiometric coefficients & \(1\) \\
        \(S_{i\alpha}\) & stoichiometric matrix element & \(1\) \\
        \(c_i\) & concentration of species \(i\) & \(\mathsf{C}\) \\
        \(c^\circ\) & standard reference concentration & \(\mathsf{C}\) \\
        \(c_i^{\mathrm{ext}}\) & reservoir concentration & \(\mathsf{C}\) \\
        \(J^\pm_\alpha\) & directional reaction flux
            & \(\mathsf{C}\mathsf{T}^{-1}\) \\
        \(J_\alpha\) & net reaction flux
            & \(\mathsf{C}\mathsf{T}^{-1}\) \\
        \(J_i^{\mathrm{ext}}\) & exchange flux
            & \(\mathsf{C}\mathsf{T}^{-1}\) \\
        \(k^\pm_\alpha\) & activity-normalized rate constant
            & \(\mathsf{C}\mathsf{T}^{-1}\) \\
        \(\gamma\) & flow rate & \(\mathsf{T}^{-1}\) \\
        \(\tau\) & reference timescale & \(\mathsf{T}\) \\
        \(\mu_i^\circ\) & standard chemical potential & \(\mathsf{E}\) \\
        \(G^\ddagger_\alpha\) & transition-state free energy
            & \(\mathsf{E}\) \\
        \(D_\alpha\) & thermodynamic drive & \(\mathsf{E}\) \\
        \(\Delta G^{\ddagger,\pm}_\alpha\)
            & effective activation free energy & \(\mathsf{E}\) \\
    \end{tabular}
    \end{ruledtabular}
\end{table}

All reactions are reversible and written as:
\begin{equation}
    \sum_{i=1}^N \nu^+_{i\alpha}X_i
    \;\rlh\;
    \sum_{i=1}^N \nu^-_{i\alpha}X_i,
    \label{eq:apx_reaction}
\end{equation}
where \(\nu^+_{i\alpha}\geq0\) and \(\nu^-_{i\alpha}\geq0\) are the
stoichiometric coefficients of species \(i\) on the two sides of reaction \(\alpha\). We orient every reaction so that the forward direction, denoted by \(+\), is ligation and the reverse direction, denoted by \(-\), is
cleavage. The inverse orientation would produce the same dynamics under proper permutation of parameters and this is just a convenient choice. The connectivity of species via reactions is defined by the stoichiometric matrix:
\begin{equation}
    S_{i\alpha}
    =
    \nu^-_{i\alpha}-\nu^+_{i\alpha}.
    \label{eq:apx_stoich}
\end{equation}

The system of Ordinary Differential Equations (ODEs) that describes the time evolution of the concentration \(c_i\) of each species is specified as
\begin{equation}
    \dv{c_i}{t}
    =
    \sum_{\alpha=1}^{R}S_{i\alpha}J_\alpha
    +
    J_i^{\mathrm{ext}},
    \label{eq:apx_ode}
\end{equation}
The first term
\begin{equation}
    J_\alpha=J^+_\alpha-J^-_\alpha
    \label{eq:apx_netflux}
\end{equation}
represents the net flux of reaction \(\alpha\) that obeys mass-action kinetics. Directional fluxes are
\begin{equation}
    J^\pm_\alpha
    =
    k^\pm_\alpha
    \prod_{i=1}^N
    \left(\frac{c_i}{c^\circ}\right)^{\nu^\pm_{i\alpha}}.
    \label{eq:apx_massaction}
\end{equation}
and dividing concentrations by the standard concentration $c^\circ$ (conventionally set to $1M$) gets all kinetic constants $k^\pm_\alpha$ measured in terms of concentration over time. 

We model environmental exchange using a Continuously Stirred Tank Reactor (CSTR). The reactor is supplied from a large reservoir while all species are removed at the same dilution rate. The corresponding exchange flux is
\begin{equation}
    J_i^{\mathrm{ext}}
    =
    \gamma\left(c_i^{\mathrm{ext}}-c_i\right),
    \label{eq:apx_cstr}
\end{equation}
where \(c_i^{\mathrm{ext}}\) is the concentration of species \(i\) in the reservoir and \(\gamma\) is the flow rate. A species with \(c_i^{\mathrm{ext}}=0\) receives no direct inflow, although it can still be produced by internal reactions.


\subsection{Physicochemical Parameterization of Rate Constants}
\label{apx:rateconstants}

We express each directional rate constant in terms of an effective activation free energy,
\begin{equation}
    k^\pm_\alpha
    =
    k_0
    \exp\left(
        -\frac{\Delta G^{\ddagger,\pm}_\alpha}{k_BT}
    \right),
    \qquad
    k_0=\frac{c^\circ}{\tau}.
    \label{eq:apx_eyring}
\end{equation}
The effective barriers are
\begin{equation}
    \Delta G^{\ddagger,\pm}_\alpha
    =
    G^\ddagger_\alpha
    -
    \sum_{i=1}^N\nu^\pm_{i\alpha}\mu_i^\circ
    \mp
    \frac{1}{2}D_\alpha.
    \label{eq:apx_barrier}
\end{equation}

The standard chemical potential \(\mu_i^\circ\) is assigned to species \(i\) and is shared by every reaction containing that species. The transition-state energy \(G^\ddagger_\alpha\) is assigned to reaction \(\alpha\). It sets the energy of the activated complex and changes the forward and reverse barriers equally. The thermodynamic drive \(D_\alpha\) is also assigned to a reaction, but enters the two directions with opposite signs. A positive \(D_\alpha\) lowers the forward barrier and raises the reverse barrier by the same amount.

The different roles of these parameters can be seen from the ratio of the
directional rate constants,
\begin{equation}
    \frac{k^+_\alpha}{k^-_\alpha}
    =
    \exp\left[
        -\frac{
            \sum_{i=1}^NS_{i\alpha}\mu_i^\circ-D_\alpha
        }{k_BT}
    \right].
    \label{eq:apx_rate_ratio}
\end{equation}
The transition-state energy cancels from this expression. The directional
bias is therefore controlled by the standard chemical potentials and the
thermodynamic drive.

Their geometric mean is
\begin{equation}
    \sqrt{k^+_\alpha k^-_\alpha}
    =
    k_0
    \exp\left[
        -\frac{
            G^\ddagger_\alpha
            -
            \frac{1}{2}
            \sum_{i=1}^N
            \left(
                \nu^+_{i\alpha}+\nu^-_{i\alpha}
            \right)\mu_i^\circ
        }{k_BT}
    \right].
    \label{eq:apx_rate_geometric_mean}
\end{equation}
The thermodynamic drive cancels from the geometric mean. Thus
\(G^\ddagger_\alpha\) changes the reaction timescale without directly changing
its directional bias. The chemical potentials enter both expressions because
they set the energies of the chemical states on either side of the reaction.

When \(D_\alpha=0\) for every reaction, the rate ratios are generated entirely
by the species-level chemical potentials and satisfy detailed balance around
every closed reaction cycle. Nonzero \(D_\alpha\) provides a nonconservative
thermodynamic drive that can break detailed balance.

\subsection{Dimensionless Formulation}
\label{apx:nondim}

The system is computationally implemented using dimensionless quantities. Concentrations are adimensionalized by normalizing over \(c^\circ\), times by \(\tau\), and energies by \(k_BT\), resulting in the dimensionless quantities below:
\begin{equation}
\begin{aligned}
    \hat{c}_i &= \frac{c_i}{c^\circ},
    &
    \hat{c}_i^{\mathrm{ext}}
        &= \frac{c_i^{\mathrm{ext}}}{c^\circ},
    &
    \hat{t} &= \frac{t}{\tau},
    \\
    \hat{\gamma} &= \gamma\tau,
    &
    \hat{\mu}_i^\circ &= \frac{\mu_i^\circ}{k_BT},
    &
    \hat{G}^\ddagger_\alpha
        &= \frac{G^\ddagger_\alpha}{k_BT},
    \\
    \hat{D}_\alpha &= \frac{D_\alpha}{k_BT},
    &
    \hat{J}^\pm_\alpha &= \frac{\tau J^\pm_\alpha}{c^\circ},
    &
    \hat{k}^\pm_\alpha
        &= \frac{\tau k^\pm_\alpha}{c^\circ}.
\end{aligned}
\label{eq:apx_rescaling}
\end{equation}

The dynamics retain their form,
\begin{equation}
    \dv{\hat{c}_i}{\hat{t}}
    =
    \sum_{\alpha=1}^{R}
    S_{i\alpha}
    \left(
        \hat{J}^+_\alpha-\hat{J}^-_\alpha
    \right)
    +
    \hat{\gamma}
    \left(
        \hat{c}_i^{\mathrm{ext}}-\hat{c}_i
    \right),
    \label{eq:apx_dimensionless_ode}
\end{equation}
with adimensional fluxes
\begin{equation}
    \hat{J}^\pm_\alpha
    =
    \hat{k}^\pm_\alpha
    \prod_{i=1}^N
    \hat{c}_i^{\,\nu^\pm_{i\alpha}}.
    \label{eq:apx_dimensionless_flux}
\end{equation}
And kinetics become
\begin{equation}
    \hat{k}^\pm_\alpha
    =
    \exp\left(
        -\Delta\hat{G}^{\ddagger,\pm}_\alpha
    \right),
    \label{eq:apx_dimensionless_rates}
\end{equation}
where energies are measured now in units of $k_BT$
\begin{equation}
    \Delta\hat{G}^{\ddagger,\pm}_\alpha
    =
    \hat{G}^\ddagger_\alpha
    -
    \sum_{i=1}^N
    \nu^\pm_{i\alpha}\hat{\mu}_i^\circ
    \mp
    \frac{1}{2}\hat{D}_\alpha.
    \label{eq:apx_dimensionless_barrier}
\end{equation}

All quantities in the remaining appendices are expressed in these dimensionless units. We therefore drop the hats from this point onward. A reported energy of \(5\) denotes \(5k_BT\), a concentration of \(1\) denotes
\(c^\circ\), and a flow rate of \(10^{-3}\) denotes
\(10^{-3}\tau^{-1}\).

\subsection{Polymerization Networks}
\label{apx:chemistry}

A polymerization chemistry is specified by a finite alphabet of monomers
$\mathcal{A}$, with $A = |\mathcal{A}|$, together with a maximum polymer
length $L$. Species are identified with nonempty words over $\mathcal{A}$
of length at most $L$, so that the species set and its cardinality are
\begin{equation}
    \mathcal{S} = \bigcup_{\ell=1}^{L}\mathcal{A}^{\ell},
    \qquad
    N = \sum_{\ell=1}^{L}A^{\ell} = \frac{A\left(A^{L}-1\right)}{A-1}.
    \label{eq:apx_species_count}
\end{equation}
Reactions are generated by cutting words. For every $s\in\mathcal{S}$ with
$|s| = \ell \geq 2$ and every cut position $1\leq j\leq \ell-1$, we write
$s = uv$ with $u = s_1\cdots s_j$ and $v = s_{j+1}\cdots s_\ell$, and
include the reversible reaction
\begin{equation}
    u + v \rlh s,
    \label{eq:apx_ligation}
\end{equation}
oriented so that ligation is the forward direction and cleavage the reverse. Both fragments are themselves words of length below $\ell\leq L$ and therefore belong to $\mathcal{S}$, so the network is closed under cleavage by construction, and no ligation can produce a word longer than $L$. When $u = v$, which occurs for even $\ell$ at $j = \ell/2$, the reaction is the dimerization $2u \rlh s$ and the reactant stoichiometric coefficient is $\nu^{+}_{u\alpha} = 2$.

Because reactants are unordered, distinct cut positions of the same word may specify the same reaction, and each such reaction is retained only once. The cuts $j$ and $\ell-j$ yield the same reactant pair precisely when $u$ and $v$ commute as words, $uv = vu$, which holds if and only if $s$ is a power of a shorter word whose length divides $\gcd(j,\ell)$.
There are exactly $A^{\gcd(j,\ell)}$ such words of length $\ell$, so after removing the redundant cuts the number of reversible reactions is
\begin{equation}
    R
    =
    \sum_{\ell=2}^{L}
    \left[
        (\ell-1)A^{\ell}
        -
        \sum_{j=1}^{\lceil \ell/2\rceil-1}
        A^{\gcd(j,\ell)}
    \right].
    \label{eq:apx_reaction_count}
\end{equation}
The first term counts all cuts and coincides with the enumeration of Ref.~\cite{Moyer2021-se}; the second removes the coincidences above.
For each choice of $\mathcal{A}$ and $L$ the network is generated deterministically and in full, and no species or reaction is sampled, pruned, or removed at any point during training.

\section{Steady-State Training and Dynamical Verification}
\label{apx:training}

Training requires evaluating the steady-state response at each point on the
input grid, differentiating the fitting objective with respect to the
physicochemical parameters, and updating these. After updating parameters, the system will result in a new steady state and the training will go over another loop on the following training step. We regularize the objective to favor steady states that are accurately resolved, dynamically stable, and numerically well conditioned. After training, we independently verify the response by integrating the full ODE dynamics.

\subsection{Training Objective}
\label{apx:trainingobjective}

For each input point \(x_k\), the steady-state concentrations of each species at index $i$ \(c^*_{i,k}\) satisfy
\begin{equation}
    F_i(c^*_{1,k},\ldots,c^*_{N,k};\theta)=0,
    \label{eq:steady_state_condition}
\end{equation}
where $F_i$ is the residual function of species $i$ that defines its time evolution
\begin{equation}
    F_i(c_1,\ldots,c_N;\theta)
    =
    \sum_{\alpha=1}^{R}S_{i\alpha}J_\alpha
    +
    \gamma(c_i^{\mathrm{ext}}-c_i).
    \label{eq:steady_state_residual}
\end{equation}
The parameter set \(\theta\) contains all the trainable values of \(G^\ddagger_\alpha\), \(\mu_i^\circ\), \(D_\alpha\) and \(\gamma\). Let \(c_y^*\) denote the readout species steady state and $f$ the target function. The CRN response at input \(x_k\) is \(c_y^*(x_k)\). For a grid of \(K\) input values, the fitting loss is
\begin{equation}
    \mathcal{L}_{\mathrm{fit}}
    =
    \frac{1}{K}
    \sum_{k=1}^{K}
    \left[c_y^*(x_k)-f(x_k)\right]^2.
    \label{eq:fitting_loss}
\end{equation}
This loss, together with the following regularization terms, is used to update $\theta$ by using a standard Adam optimizer~\cite{Kingma2014-xh}.

\subsubsection{Regularization terms}
\label{apx:residualregularization}
We employ three main regularization terms, aimed at ensuring the training procedure stays in parameter regions where steady state solutions~\ref{apx:ptc} are of good quality, stable and non-singular. We do this via residual, spectral and conditioning regularization respectively. When evaluating and differentiating parameters against regularization terms we keep steady-state concentrations fixed.

Residual regularization ensures that numerical steady-state solutions have low residuals, which correspond to high quality steady state solutions being truly near a zero of the residual function. We penalize high residual using the following regularization term 
\begin{equation}
    \mathcal{L}_{\mathrm{res}}
    =
    \frac{1}{K}
    \sum_{k=1}^{K}
    \sum_{i=1}^{N}
    F_i(\bar c^*_{1,k},\ldots,\bar c^*_{N,k};\theta)^2,
    \label{eq:residual_regularization}
\end{equation}
where
\begin{equation}
    \bar c^*_{i,k}
    =
    \texttt{stop\_gradient}(c^*_{i,k}).
\end{equation}

To ensure stability we regularize agains the eigenvalues of the Jacobian of the ODE system. A root of \(F_i=0\) is dynamically stable only if every eigenvalue of the dynamical Jacobian $\mathbb{J}$ has a negative real part. At each input point, we evaluate
\begin{equation}
    \mathbb{J}_{ij}^{(k)}
    =
    \left.
    \pdv{F_i}{c_j}
    \right|_{\bar c^*_{1,k},\ldots,\bar c^*_{N,k}},
    \label{eq:dynamical_jacobian}
\end{equation}
and define its spectral abscissa as
\begin{equation}
    s_k
    =
    \max_j \operatorname{Re}\zeta_{j,k},
    \label{eq:spectral_abscissa}
\end{equation}
where \(\zeta_{j,k}\) are the eigenvalues of \(\mathbb{J}_{ij}^{(k)}\). The spectral
penalty is
\begin{equation}
    \mathcal{L}_{\mathrm{eig}}
    =
    \frac{1}{K}
    \sum_{k=1}^{K}
    \operatorname{ReLU}(s_k)^2.
    \label{eq:spectral_regularization}
\end{equation}
This penalty discourages roots with eigenvalues in the right half-plane. Though, it does not by itself guarantee that the trained response belongs to a single dynamically accessible steady-state branch, which is a condition we check post-training via quasistatic verification~\ref{apx:odecheck}.

A nearly singular dynamical Jacobian makes both root finding and implicit differentiation unreliable. We regularize against this by measuring its conditioning
\begin{equation}
    \chi_k
    =
    \log_{10}
    \left(
        \frac{\sigma_{\max,k}}{\sigma_{\min,k}}
    \right),
    \label{eq:log_condition_number}
\end{equation}
where \(\sigma_{\max,k}\) and \(\sigma_{\min,k}\) are the largest and smallest singular values of \(\mathbb{J}_{ij}^{(k)}\). The conditioning penalty is
\begin{equation}
    \mathcal{L}_{\mathrm{cond}}
    =
    \frac{1}{K}
    \sum_{k=1}^{K}
    \operatorname{ReLU}
    \left(
        \chi_k-\chi_{\mathrm{thresh}}
    \right)^2,
    \label{eq:conditioning_regularization}
\end{equation}
with \(\chi_{\mathrm{thresh}}=9\), corresponding to a condition-number threshold of \(10^9\). As in the residual and spectral penalties, the steady-state concentrations are held fixed during differentiation.

The complete training objective is
\begin{equation}
    \mathcal{L}
    =
    \mathcal{L}_{\mathrm{fit}}
    +\lambda_{\mathrm{res}}\mathcal{L}_{\mathrm{res}}
    +\lambda_{\mathrm{eig}}\mathcal{L}_{\mathrm{eig}}
    + \lambda_{\mathrm{cond}}\mathcal{L}_{\mathrm{cond}}
\label{eq:total_training_loss}
\end{equation}

The fitting-loss gradient accounts for the dependence of the steady state on the parameters and the regularization gradients are evaluated while holding the numerical steady-state solutions fixed.
We found the following regularization coefficients to be optimal and kept them fixed during all our experiments to: $\lambda_{\mathrm{res}}=10$, $\lambda_{\mathrm{eig}}=10^{-2}$ and $\lambda_{\mathrm{cond}}=0.4$. These values however may be specific for the target function range and amplitude values we used in our experiments and may need to be tuned for different experimental settings. 

\subsection{Steady States by Pseudo-Transient Continuation}
\label{apx:ptc}

Every training epoch requires finding the current response solution to compute the loss from. Rather than integrating the often stiff dynamics to steady state during every training epoch, we solve \(F_i=0\) using pseudo-transient continuation. This method follows a sequence of implicit pseudo-time steps that approaches a Newton iteration near a root. At iteration \(n\), we solve
\begin{equation}
    \sum_{j=1}^{N}
    \left(
        \frac{\delta_{ij}}{\Delta t_n}
        -
        \mathbb{J}_{ij}^{(n)}
    \right)
    \Delta c_j^{(n)}
    =
    F_i(c_1^{(n)},\ldots,c_N^{(n)};\theta),
    \label{eq:ptc_linear_system}
\end{equation}
where
\begin{equation}
    \mathbb{J}_{ij}^{(n)}
    =
    \left.
    \pdv{F_i}{c_j}
    \right|_{c_1^{(n)},\ldots,c_N^{(n)}}.
\end{equation}
The concentrations are updated according to
\begin{equation}
    c_i^{(n+1)}
    =
    \max\left(
        c_i^{(n)}+\Delta c_i^{(n)},\,
        \epsilon_c
    \right),
    \label{eq:ptc_update}
\end{equation}
which prevents the numerical iterate from entering the negative concentration domain.

For small \(\Delta t_n\), the diagonal term in Eq.~\eqref{eq:ptc_linear_system} limits the size of the update. As \(\Delta t_n\) increases, the method approaches a full Newton step. We adapt the pseudo-time step using switched evolution relaxation,
\begin{equation}
\begin{aligned}
    \Delta t_{n+1}
    =
    \min\Bigg(
        \Delta t_0\frac{r_0}{r_{n+1}},
        g\,\Delta t_n,\,
        \Delta t_{\max}
    \Bigg),
\end{aligned}
\label{eq:ser_update}
\end{equation}
where
\begin{equation}
    r_n
    =
    \max_i
    \left|
        F_i(c_1^{(n)},\ldots,c_N^{(n)};\theta)
    \right|.
\end{equation}
The factor \(g\) limits the growth of the pseudo-time step. Convergence is declared when
\begin{equation}
    r_n
    <
    \epsilon_{\mathrm{abs}}
    +
    \epsilon_{\mathrm{rel}}r_0.
    \label{eq:ptc_convergence}
\end{equation}

Pseudo-transient continuation follows the local dynamics more closely than an unconstrained Newton solve and can therefore favor stable roots. At the first training step, the procedure is initialized to a concentration vector of all zeroes, while for each subsequent training step step $t$, the procedure starts from an initial guess equivalent to the steady solution at the previous $t-1$ training step.

\subsection{Implicit Differentiation and Adjoint Gradients}
\label{apx:implicitgradients}

To update parameters on the fit loss $\mathcal{L}_{\mathrm{fit}}$, differentiating through every iteration of the root finder would require storing and backpropagating through its full solution path, which would be computationally costly. Instead, we differentiate the steady-state solution directly to parameters by leveraging the implicit function theorem~\cite{Blondel2021-wr}.

For a trainable parameter \(\theta_p\), differentiating \(F_i=0\) gives
\begin{equation}
    \sum_{j=1}^{N}
    \mathbb{J}_{ij}
    \pdv{c_j^*}{\theta_p}
    +
    \pdv{F_i}{\theta_p}
    =
    0.
    \label{eq:implicit_derivative}
\end{equation}
We avoid computing \(\partial c_j^*/\partial\theta_p\) explicitly by solving the adjoint system (where $\lambda_i$ is the adjoint variable, not to be confused with regularization coefficients)
\begin{equation}
    \sum_{i=1}^{N}
    \mathbb{J}_{ij}\lambda_i
    =
    \pdv{\mathcal{L}_{\mathrm{fit}}}{c_j^*}.
    \label{eq:adjoint_system}
\end{equation}
The fitting-loss gradient then follows from
\begin{equation}
    \pdv{\mathcal{L}_{\mathrm{fit}}}{\theta_p}
    =
    -
    \sum_{i=1}^{N}
    \lambda_i
    \pdv{F_i}{\theta_p}.
    \label{eq:adjoint_parameter_gradient}
\end{equation}
An independent adjoint solve is performed at each input point. The resulting parameter gradients are combined across the input grid.

\subsection{Trusted Gradients and Parameter Updates}
\label{apx:parameterupdates}

A grid point contributes to the implicit gradient only when its root-finding
residual satisfies
\begin{equation}
    \max_i
    \left|
        F_i(c^*_{1,k},\ldots,c^*_{N,k};\theta)
    \right|
    \leq
    \epsilon_{\mathrm{trust}}.
    \label{eq:trusted_root}
\end{equation}
This trust mask prevents poorly resolved roots from producing unreliable
adjoint gradients. If more than \(50\%\) of the input points are untrusted,
the update is discarded and training terminates with a
\texttt{solver\_unstable} exit condition.

Before each parameter update, the gradient associated with each trainable
parameter class is clipped by its \(\ell_2\) norm,
\begin{equation}
    g_p^{\mathrm{clip}}
    =
    g_p
    \min\left(
        1,\,
        \frac{g_{\max}}{\lVert g\rVert_2}
    \right),
    \label{eq:gradient_clipping}
\end{equation}
where the norm is evaluated over all parameters in the same class and
\(g_{\max}=10\).

\subsection{Stopping Conditions and Exit Labels}
\label{apx:stopping}

Training terminates when the first of the following conditions is met:
\begin{enumerate}
    \item The total gradient norm falls below \(2\times10^{-4}\).
    \item The relative change in loss over 500 epochs falls below \(10^{-6}\).
    \item More than \(50\%\) of the input points fail the trusted-root
    criterion in Eq.~\eqref{eq:trusted_root}.
    \item The run reaches the maximum budget of \(16{,}000\) epochs.
\end{enumerate}

These exit conditions do not determine whether a run is counted as successful, which additionally depends on its fit quality and post-training dynamical verification.

\subsection{Quasistatic ODE Verification}
\label{apx:odecheck}

After training, we verify that the predicted response can be reached by the actual dynamics of the system. We numerically integrate
\begin{equation}
    \dv{c_i}{t}=F_i(c_1,\ldots,c_N;\theta)
\end{equation}
using the implicit Kvaerno5 solver with adaptive step-size control \cite{kidger2021on}. We use \(\mathrm{atol}=10^{-6}\), \(\mathrm{rtol}=10^{-6}\), and a maximum integration time of \(5\times10^6\). 

Verification proceeds quasistatically along the ordered input grid. The first input point is initialized from its trained root-finding solution. Each subsequent point is initialized from the converged ODE state at the preceding grid point. This procedure tests whether the fitted response lies on a dynamically reachable branch that can be followed as the input changes.

A point is considered converged when
\begin{equation}
    \max_i|F_i|
    \leq
    10^{-4}.
    \label{eq:ode_convergence}
\end{equation}
If integration doesn't converge at a point for more than a hard time budget of $180$ seconds, we flag the point as failed and continue onward.

Failing even one grid point discards a run as failed, as runs pass dynamical verification only if all \(K\) input points converge and the ODE readout agrees with the root-finding readout according to
\begin{equation}
    \left[
        \frac{1}{K}
        \sum_{k=1}^{K}
        \left(
            y_{\mathrm{ODE}}(x_k)
            -
            y^*(x_k)
        \right)^2
    \right]^{1/2}
    <
    10^{-2}.
    \label{eq:ode_verification_rmse}
\end{equation}
This final test rejects responses that depend on unstable roots, disconnected
steady-state branches, or trajectories that do not settle to the predicted
fixed points.

\section{Target Ensemble}
\label{apx:targets}

The scaling and parameter-freezing studies use random squared trigonometric polynomials as target functions. The construction gives smooth, strictly positive profiles on a fixed output window, with a single integer knob \(H\) that sets how many oscillations the target can develop.

The reactor input \(x\in[x_{\min},x_{\max}]\) is first mapped linearly onto an angular coordinate,
\begin{equation}
    t(x) = 2\pi\,\frac{x-x_{\min}}{x_{\max}-x_{\min}},
    \label{eq:target_input_mapping}
\end{equation}
so that the input range covers exactly one period. For each target we draw i.i.d.\ complex coefficients \(a_\omega=u_\omega+\mathrm{i}v_\omega\) with \(u_\omega,v_\omega\sim\mathcal{N}(0,1)\) for \(\omega=0,\ldots,H\), and take the squared modulus of the resulting harmonic sum,
\begin{equation}
    q(x) = \left|\sum_{\omega=0}^{H}a_\omega\,e^{\mathrm{i}\omega t(x)}\right|^{2}.
    \label{eq:target_raw}
\end{equation}
Squaring makes \(q\) nonnegative everywhere and therefore admissible as a concentration readout.

The draw fixes the shape of the target but not its scale, so we rescale \(q\) onto the output window \([f_{\min},f_{\max}]\). Writing \(q_{\min}\) and \(q_{\max}\) for the extrema of \(q\) over the sampled input grid, the target is
\begin{align}
    f(x) 
    &= \bar f + \alpha\,(f_{\max}-f_{\min})
    \left[\frac{q(x)-q_{\min}}{q_{\max}-q_{\min}}-\frac{1}{2}\right] \\
    \qquad \bar f 
    &= \frac{f_{\min}+f_{\max}}{2}.
    \label{eq:target_final}
\end{align}
The bracket lies in \([-\tfrac12,\tfrac12]\) by construction, so \(\alpha\) sets the amplitude about the midpoint \(\bar f\): the target is constant at \(\alpha=0\) and spans the full window at \(\alpha=1\).

Independent draws at the same \(H\) give distinct targets of comparable spectral complexity, and increasing \(H\) admits more peaks, valleys, and changes in curvature. We therefore use \(H\) as our measure of target complexity throughout. Throughout our experiments we always set $f_{\min} = 0.2$ and $f_{\max} = 0.6$.

\section{Illustrative Fits}
\label{apx:examples}

The examples in Fig.~\ref{fig:fig1}(D--F) demonstrate the range of responses that can be obtained with the training procedure described in Appendices~\ref{apx:crn} and~\ref{apx:training}. The input is the reservoir concentration of the designated input monomer, and the output is the steady-state concentration of the designated readout species.

Figure~\ref{fig:fig1}(D) shows the training trajectory of an \(abcdef3\) network with $N=258$ species and $R=462$ reactions, against a trigonometric polynomial target (Appendix.\ref{apx:targets}) with $H=5$ harmonics, 
the response curves shown at intermediate epochs $800$, $8000$ and $12000$ are evaluated from the steady-state solutions stored during training. The input species is the monomer $a$ and auxiliary monomers $b,c,d,e,f$ are provided to the system at a fixed reservoir concentration of $1$. The output species is the dimer $ab$.

Figure~\ref{fig:fig1}(E) contains examples trained against smooth target functions chosen to illustrate qualitatively different response shapes, including localized peaks and sharp transitions. Every target function is sampled on a grid of $32$ points. From top left to bottom right:
\begin{itemize}
    \item Target: Gaussian $a\exp\big({-{1 \over 2}{(x-\mu)^2\over \sigma ^2}}\big)$ with $a = 0.6$, $\mu=3$, $\sigma = 0.7$. Topology: $ab4$. 
    \item Target: Step $f(x) = f_{\min} + \frac{f_{\max} - f_{\min}}{1 + \exp({-s(x - x_0)})}$ where $f
    _{\min} = 0.02$, $f_{\max}=0.6$, $s = 10$, $x_0 = 3$. Topology: $ab5$.
    \item Target: a random Legendre polynomial (rescaled) $$\tilde{f}(t) = \sum_{k=0}^{5} c_k \, L_k(t), \quad t = \frac{2(x - x_{\min})}{x_{\max} - x_{\min}} - 1$$
    $$f(x) = \frac{a \bigl(\tilde{f}(t) - \tilde{f}_{\min} + b\bigr)}{\tilde{f}_{\max} - \tilde{f}_{\min} + b}$$
    where coefficients are sampled from the normal distribution $c_k \sim \mathcal{N}(0,1)$ with seed 3, $a = 0.6$ (amp), $b = 0.2$ (floor), and $L_k$ are Legendre polynomials. Topology: $ab4$.
    \item Target: a random trigonometric polynomial \ref{apx:targets} with $6$ harmonics. Topology: $ab6$
    \item Target: a random trigonometric polynomial \ref{apx:targets} with $8$ harmonics. Topology: $ab6$ 
    \item Target: von Neumann elephant \ref{apx:multireadout}. Topology: $ab6$
\end{itemize}

Figure~\ref{fig:fig1}(F) compares \(ab3\) and \(ab4\) against the same quadratic tent target. $$f(x) = \min\left(\, a\,t_L(x)^2,\;\; a\,t_R(x)^2 \,\right)$$
  $$t_L(x) = \frac{x - (x_0 - w)}{w}, \quad t_R(x) = \frac{(x_0 + w) - x}{w}$$
  where $x_0 = 3.0$, $w = 2.0$ (half-width to edge), $a = 0.6$. The input $a$ and output $ab$ species are held fixed so that the comparison isolates the effect of increasing the maximum polymer length.

\subsection{Multiple-Readout Training}
\label{apx:multireadout}

The parametric example in Fig.~\ref{fig:fig1}(E) uses one input and two readout species. Let \(c^*_{1,k}\) and \(c^*_{2,k}\)  denote the steady state concentration of readouts at the $k$th grid point, with target coordinates \(f_1(x)\) and \(f_2(x)\). We train both coordinates jointly using the loss
\begin{equation}
  \mathcal{L}_{\mathrm{fit}}
  =
  \frac{1}{2K}
  \sum_{k=1}^{K}
  \sum_{l=1}^{2}
  \left[
      c^*_{l,k}-f_l(x_k)
  \right]^2.
  \label{eq:multiple_readout_loss}
\end{equation}

For the von Neumann elephant, the network is $ab6$ (126 species, 502 reactions). The inflow concentration of
the dimer $ab$ serves as the input $x\in[1,5]$, sampled at $K=64$ uniformly spaced points, while the monomers $a$ and $b$ are held fixed at unit concentration. The two readout species are $aba$ and $bab$.

The target follows the Fourier parametrization of \cite{Mayer2010-if}:
\begin{align*}
  \tilde{f}_1(\phi) &= -60\cos\phi - 30\sin\phi + 8\sin 2\phi - 10\sin 3\phi \\
  \tilde{f}_2(\phi) &= -50\sin\phi - 18\sin 2\phi - 12\cos 3\phi + 14\cos 5\phi
\end{align*}
where $\phi = 2\pi(x - x_{\min})/(x_{\max}-x_{\min})$, then rescaled to the interval $[f_{\min}, f_{\max}] = [0.2,\, 0.6]$:
\begin{equation}
  f_l(x) = f_{\min} + \frac{\tilde{f}_l - \min\tilde{f}_l}
  {\max\tilde{f}_l - \min\tilde{f}_l}\,(f_{\max} - f_{\min}).
\end{equation}

\section{Scaling Study}
\label{apx:scaling}

The scaling study measures how the probability of fitting a target changes with target complexity and network size. All networks are trained using the same input and readout species, target construction, success criterion, and numerical procedure.

\subsection{Topology Set and Network-Size Measures}
\label{apx:scalingtopologies}

We include every polymerization network in the selected alphabet size and maximum polymer length range containing no more than approximately \(500\) reactions. This procedure gives the nine topologies used in Fig.~\ref{fig:fig2}, listed in Table.\ref{tab:architectures}.
The reaction networks explored in this work are generated by sweeping through monomer alphabet $A$ and maximum polymer length $L$:

\begin{itemize}
  \item \textbf{Fixed alphabet ($A = 2$), increasing polymer length.}
    \begin{itemize}
      \item $L = 3$: $N = 14$ species, $R = 18$ reactions.
      \item $L = 4$: $N = 30$ species, $R = 64$ reactions.
      \item $L = 5$: $N = 62$ species, $R = 188$ reactions.
      \item $L = 6$: $N = 126$ species, $R = 502$ reactions.
    \end{itemize}

  \item \textbf{Fixed polymer length ($L = 3$), increasing alphabet.}
    \begin{itemize}
      \item $A = 3$: $N = 39$ species, $R = 60$ reactions.
      \item $A = 4$: $N = 84$ species, $R = 140$ reactions.
      \item $A = 5$: $N = 155$ species, $R = 270$ reactions.
      \item $A = 6$: $N = 258$ species, $R = 462$ reactions.
    \end{itemize}

  \item \textbf{Off-diagonal case.} A single intermediate configuration with
    $A = 3$, $L = 4$ yields $N = 120$ species, $R = 300$ reactions.
\end{itemize}

For each topology, the number of trainable parameters simply scales as number of species $N$ (one standard chemical potential per species), $2R$ (a transition state energy and a thermodynamical drive per reaction) plus the flow rate, namely:
\begin{equation}
    P = N+2R+1.
    \label{eq:scaling_parameter_count}
\end{equation}


\begin{table}[t]
  \centering
  \caption{Polymerization topologies used in the scaling study. Each network is
  specified by its alphabet size $A$ and maximum polymer length $L$, which
  determine the number of species $N$ and reversible reactions $R$. Marker and
  color match Fig.~\ref{fig:fig2}.}
  \label{tab:architectures}
  \begin{tabular}{@{}l r r r r@{}}
    \toprule
     & $A$ & $L$ & $R$ & $N$ \\
    \midrule
    \key{bluel}{circle}{1.1pt}                      & 2 & 3 & 18  & 14  \\
    \key{bluem}{circle}{1.3pt}                      & 2 & 4 & 64  & 30  \\
    \key{blued}{circle}{1.6pt}                      & 2 & 5 & 188 & 62  \\
    \key{bluex}{circle}{1.9pt}                      & 2 & 6 & 502 & 126 \\
    \key{bluel}{regular polygon, regular polygon sides=3}{1.1pt} & 3 & 3 & 60  & 39  \\
    \key{bluem}{regular polygon, regular polygon sides=3}{1.1pt} & 3 & 4 & 300 & 120 \\
    \key{bluel}{regular polygon, regular polygon sides=4}{1.1pt} & 4 & 3 & 140 & 84  \\
    \key[5.2pt]{bluel}{regular polygon, regular polygon sides=5}{1.1pt} & 5 & 3 & 270 & 155 \\
    \key[5.6pt]{bluel}{regular polygon, regular polygon sides=6}{1.1pt} & 6 & 3 & 462 & 258 \\
    \bottomrule
  \end{tabular}
\end{table}

Increasing either alphabet size or maximum polymer length increases all three
size measures. The available polymerization topologies therefore do not vary
\(N\), \(R\), and \(P\) independently. Their strong covariation prevents the present experiment from identifying one of these quantities as the unique structural determinant of expressivity.

\subsection{Experimental Design}
\label{apx:scalingprotocol}

For every topology, the input is the reservoir concentration \(x=c_a^{\mathrm{ext}}\) of the monomer $a$, and the readout is the steady-state concentration of \(ab\). The reservoir concentrations of the remaining monomers are fixed at \(1\), and the reservoir concentrations of all polymers are zero.

The input is sampled at \(K=32\) uniformly spaced points over \([x_{\min},x_{\max}]=[1.0, 5.0]\). All three energetic parameter classes and the flow rate \(\gamma\) are
trainable.

For each topology and target complexity \(H\), we train the network against target functions drawn from the ensemble in
Appendix~\ref{apx:targets}. The experiment contains ten values of \(H\) ranging from $H=1$ to $H=10$. For each value of $H$ we sample $50$ random targets and reuse them across all topologies. This gives a total of $10\times 50$ targets that $9$ topologies are all trained against, resulting in the total training rounds launched being
\begin{equation}
    9 \times 10 \times 50 = 4500
\end{equation}

The energetic parameters and flow rate are initialized independently for each
run, i.i.d. Gaussian sampled with $\sigma^2=0.1$ and centered at the origin with mean $0$, using a unique seed for each experiment. Each network is then trained and verified using the common procedure in Appendix~\ref{apx:training}.

\subsection{Success Rate and Capacity Estimation}
\label{apx:capacityestimation}

We quantify fit quality using the coefficient of determination
\begin{equation}
  R^2
  =
  1-
  \frac{
      \sum_{k=1}^{K}
      \left[
          c^*_{y,k}-f(x_k)
      \right]^2
  }{
      \sum_{k=1}^{K}
      \left[
          f(x_k)-\overline{f}
      \right]^2
  }
  \label{eq:scaling_r2}
\end{equation}
where $c^*_{y,k}$ is the readout steady state concentration and $\overline{f}$ is the target function average over the grid
\begin{equation}
  \overline{f}
  =
  \frac{1}{K}
  \sum_{k=1}^{K}f_H(x_k)
\end{equation}

A training run is counted as successful when it satisfies both of the following conditions:
\begin{enumerate}
  \item The response passes the quasistatic ODE verification in Appendix~\ref{apx:odecheck}.
  \item The ODE verified response reaches \(R^2\geq0.98\).
\end{enumerate}

For topology \(q\) and target complexity \(H\), the empirical success rate is
\begin{equation}
  p_q(H)
  =
  \frac{1}{50}
  \sum_{r=1}^{50}I_{q,H,r}
  \label{eq:scaling_success_rate}
\end{equation}
where \(I_{q,H,r}=1\) for a successful run and \(0\) otherwise.

We define the expressivity \(H_{50}\) as the target complexity at which the success rate crosses \(50\%\). Suppose the crossing
lies between adjacent values \(H_j\) and \(H_{j+1}\). Linear interpolation gives
\begin{equation}
\begin{aligned}
  H_{50}
  ={}&
  H_j
  +
  \frac{0.5-p_q(H_j)}
  {p_q(H_{j+1})-p_q(H_j)}
  \\
  &\times
  \left(H_{j+1}-H_j\right)
\end{aligned}
\label{eq:h50_interpolation}
\end{equation}
In case an empirical success curve is not monotonic, the crossing is selected
by taking the highest \(H\) at which \(p_q(H)\geq 0.5\) and interpolating from
there, yielding the largest (most generous) \(H_{50}\), although we do not
observe nonmonotonic crossings in our experiments. If a topology does not
cross \(50\%\) at any target \(H\) we set \(H_{50}=0\), although this case also
does not arise in our results.

At each \(H\), the binomial standard error of the success rate is
\begin{equation}
  \sigma_p(H)
  =
  \sqrt{
      \frac{
          p_q(H)\left[1-p_q(H)\right]
      }{50}
  }
  \label{eq:success_rate_error}
\end{equation}
The uncertainty on \(H_{50}\) in Fig.~\ref{fig:fig2} is obtained by
propagating the standard errors of the two bracketing success rates through
the linear interpolant~\eqref{eq:h50_interpolation}:
\begin{equation}
  \sigma_{H_{50}}
  =
  \sqrt{
      \left(\frac{\partial H_{50}}{\partial p_j}\right)^2\sigma_{p_j}^2 +
      \left(\frac{\partial H_{50}}{\partial p_{j+1}}\right)^2\sigma_{p_{j+1}}^2
  }
  \label{eq:h50_error}
\end{equation}
with
\begin{align}
  \frac{\partial H_{50}}{\partial p_j}
  =&
  \frac{(H_{j+1}-H_j)(0.5-p_{j+1})}{(p_{j+1}-p_j)^2}
  \quad \\
  \frac{\partial H_{50}}{\partial p_{j+1}}
  =&
  \frac{(H_{j+1}-H_j)(p_j-0.5)}{(p_{j+1}-p_j)^2}
\end{align}

Associations between \(H_{50}\) and the network-size measures are quantified
using Spearman's rank correlation coefficient. The dashed lines in
Fig.~\ref{fig:fig2}(B--D) are unweighted ordinary least-squares fits of
\(H_{50}\) against the base-10 logarithm of each size measure.

\subsection{Failure Modes}
\label{apx:scalingfailures}

An unsuccessful run may fail during optimization, fail to reach the required fit quality, or fail the post-training dynamical verification. We record these
outcomes separately. A solver failure occurs when training exits early due to numerical instabilities as defined in Appendix~\ref{apx:stopping}. A verification failure happens when a response found during training doesn't match with the actual quasistatic response behavior exhibited by the system post-training. 

\begin{figure*}[t]
    \centering
    \includegraphics[width=\linewidth]{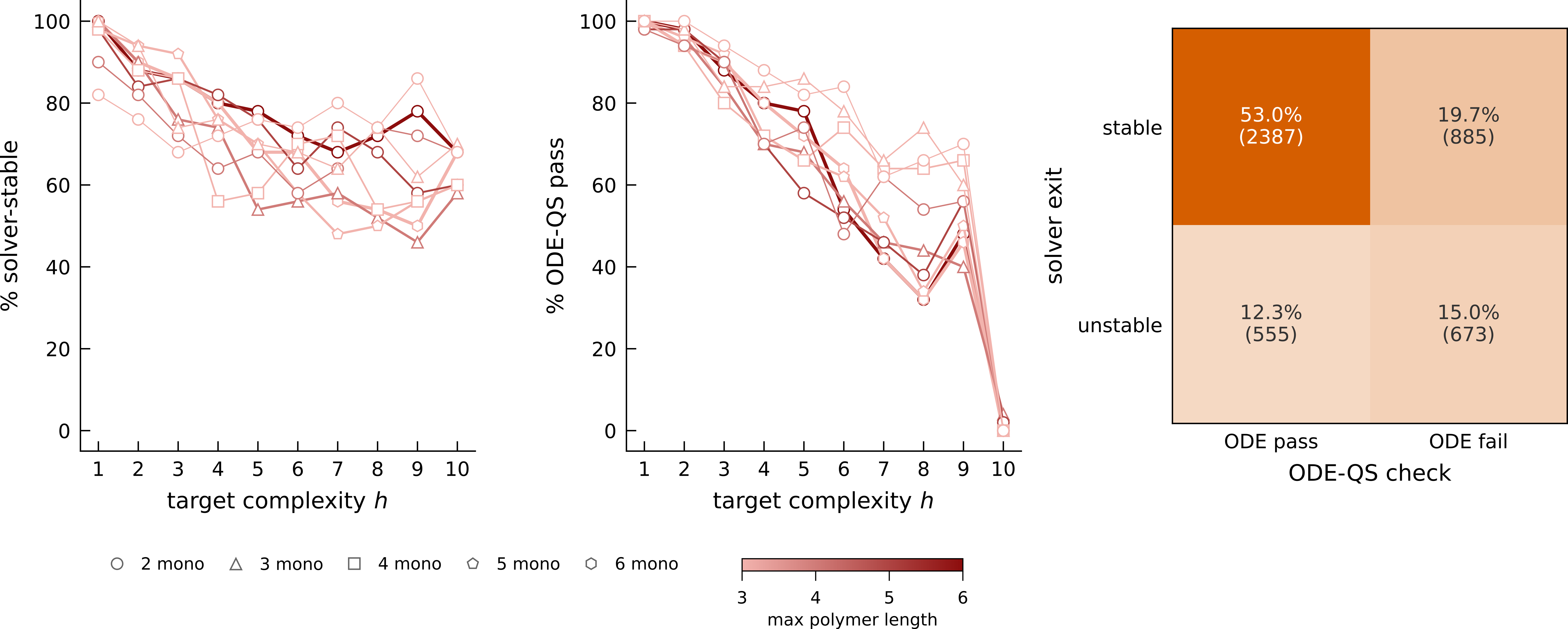}
    \caption{
    Failure statistics for the scaling study in Fig.~\ref{fig:fig2}. The left panel describes the fraction of experiments that resulted in stable numerical behavior during the training procedure, normalized by the $50$ experiments run per target complexity per topology. The central panel shows the corresponding fraction of experiments that passed the ODE quasistatic check (QS-pass). The right panel shows the distribution among $4$ possible resulting combinations of numerically stable and unstable experiments that resulted in passing or not passing the quasistatic check. We see here that it is possible that some experiments pass the quasistatic check even when they resulted in early training exit due to numerical instabilities.  
    }
    \label{fig:failure_stats}
\end{figure*}

Figure~\ref{fig:failure_stats} reports the frequency of the different outcomes
across topologies against the target complexity classes of Sec.~\ref{sec:scaling}. This analysis distinguishes failures of the optimization
procedure from fitted responses that cannot be followed by the nonlinear ODE
dynamics.

\subsection{Non-trivial Effect of Readout Choice}
\label{apx:readoutloss}

The scaling study uses the same readout species for every topology because fit quality can depend on the location of the readout within the network. We
evaluate this dependence using a fixed \(abcd3\) topology and one target drawn at \(H=3\). The reservoir concentration of \(a\) remains the input. Each of the $80$ available non-monomer species is then used in turn as the readout, while the topology, target, input grid, and training procedure are all held fixed. We run a single training round for each, resulting in a total of $80$ experiments.

\begin{figure}[t]
    \centering
    \includegraphics[width=\linewidth]{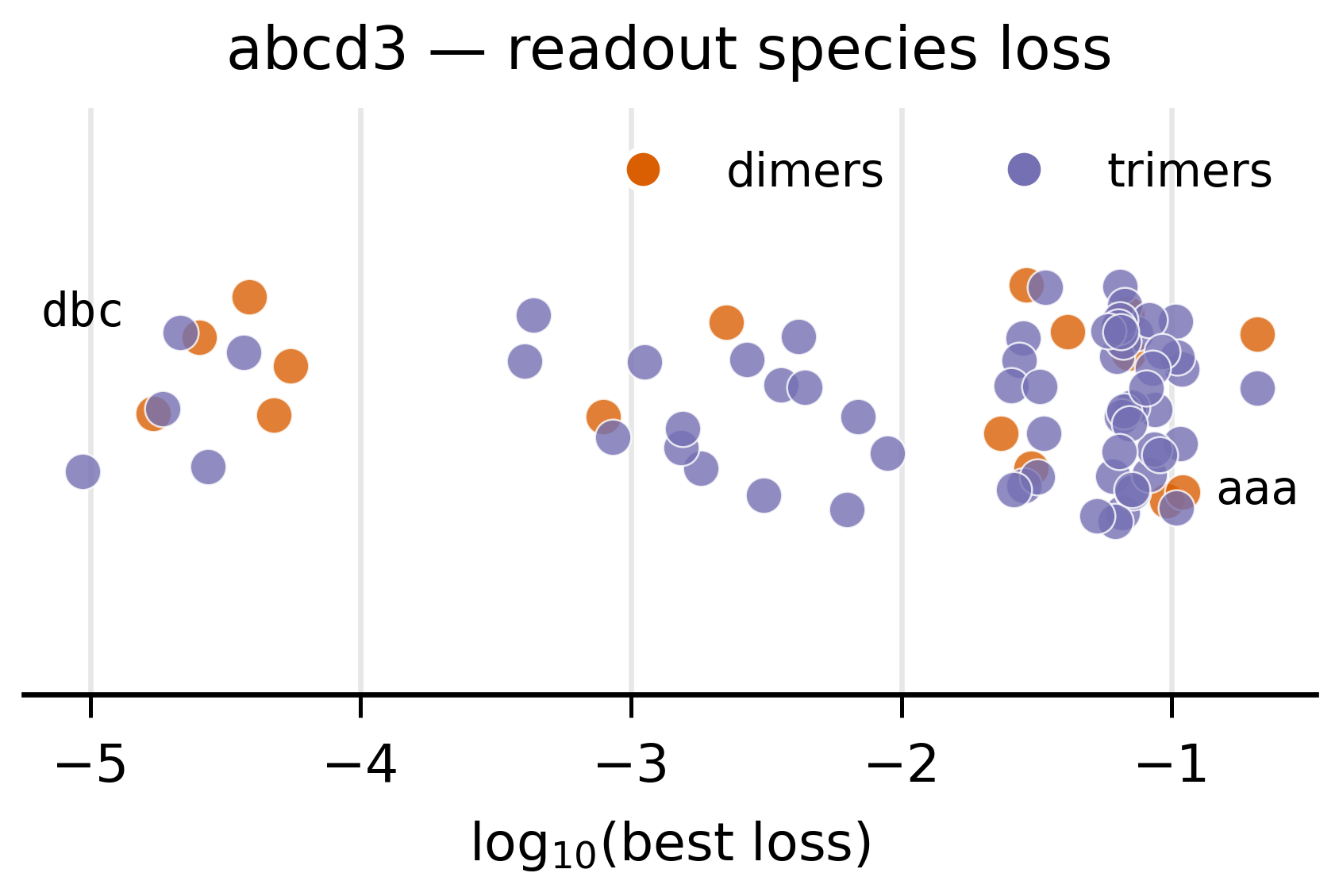}
    \caption{
    Swarm plot of the resulting fit quality on the readout species in an \(abcd3\) network trained against a fixed \(H=3\) target. Each dot corresponds to a single run on a single readout choice, its horizontal position determined by the base 10 logarithm of the best loss achieved in that training round. This highlights a non-trivial dependency on the choice of readout species we do not further investigate in the present study.  
    }
    \label{fig:readouts_loss}
\end{figure}

The variation across readouts shows that network topology alone does not determine fit quality. Fixing \(ab\) as the readout throughout the scaling study removes this source of variation from comparisons among topologies. We did not explore the effects of changing input species while keeping the readout species fixed.

\section{Parameter-Freezing Study}
\label{apx:freezing}

The parameter-freezing study tests how each physicochemical parameter class
contributes to trainability. The network topology, target ensemble, input,
readout, and flow rate are held fixed while the trainable energetic parameters
are varied.

We consider the three energetic parameter classes
\(\{G^\ddagger_\alpha\}\), \(\{\mu_i^\circ\}\), and \(\{D_\alpha\}\). The
experimental bins include each class alone, each pair of classes, and the fully trainable combination:
\textbf{1.}~$\{G^\ddagger_\alpha\}$ 
\textbf{2.}~$\{\mu_i^\circ\}$ 
\textbf{3.}~$\{D_\alpha\}$ 
\textbf{4.}~$\{G^\ddagger_\alpha,\mu_i^\circ\}$
\textbf{5.}~$\{G^\ddagger_\alpha,D_\alpha\}$ 
\textbf{6.}~$\{\mu_i^\circ,D_\alpha\}$
\textbf{7.}~$\{G^\ddagger_\alpha,\mu_i^\circ,D_\alpha\}$

A parameter class excluded from a bin is initialized to zero and is frozen during training. Trainable parameters are randomly initialized as $\theta_i\sim\mathcal{N}(0,0.1)$. The flow rate \(\gamma\) is fixed in every bin at $\gamma = 1$.

All freezing experiments use the \(abc3\) topology with species \(a\) as the input and species \(ab\) as the readout. The reservoir concentrations of the remaining monomers are fixed at \(1\), and those of all polymers are zero.

For each bin we train against \(128\) trigonometric polynomial targets with two harmonics \(H=2\) (Appendix~\ref{apx:targets}). The same targets and are reused across all bins.

After initialization, each target and parameter bin $b$ is trained using the procedure in Appendix~\ref{apx:training}. The success criterion is the same as in the scaling study. A run must reach \(R^2\geq0.98\) and pass the quasistatic ODE verification. The success rate for parameter bin \(b\) is
\begin{equation}
    p_b
    =
    \frac{1}{128}
    \sum_{r=1}^{128}I_{b,r},
    \label{eq:freezing_success_rate}
\end{equation}
since one run is performed for each target. Here \(I_{b,r}=1\) if the run is successful and \(0\) otherwise.

\begin{figure*}[t]
    \centering
    \includegraphics[width=0.6\linewidth]
    {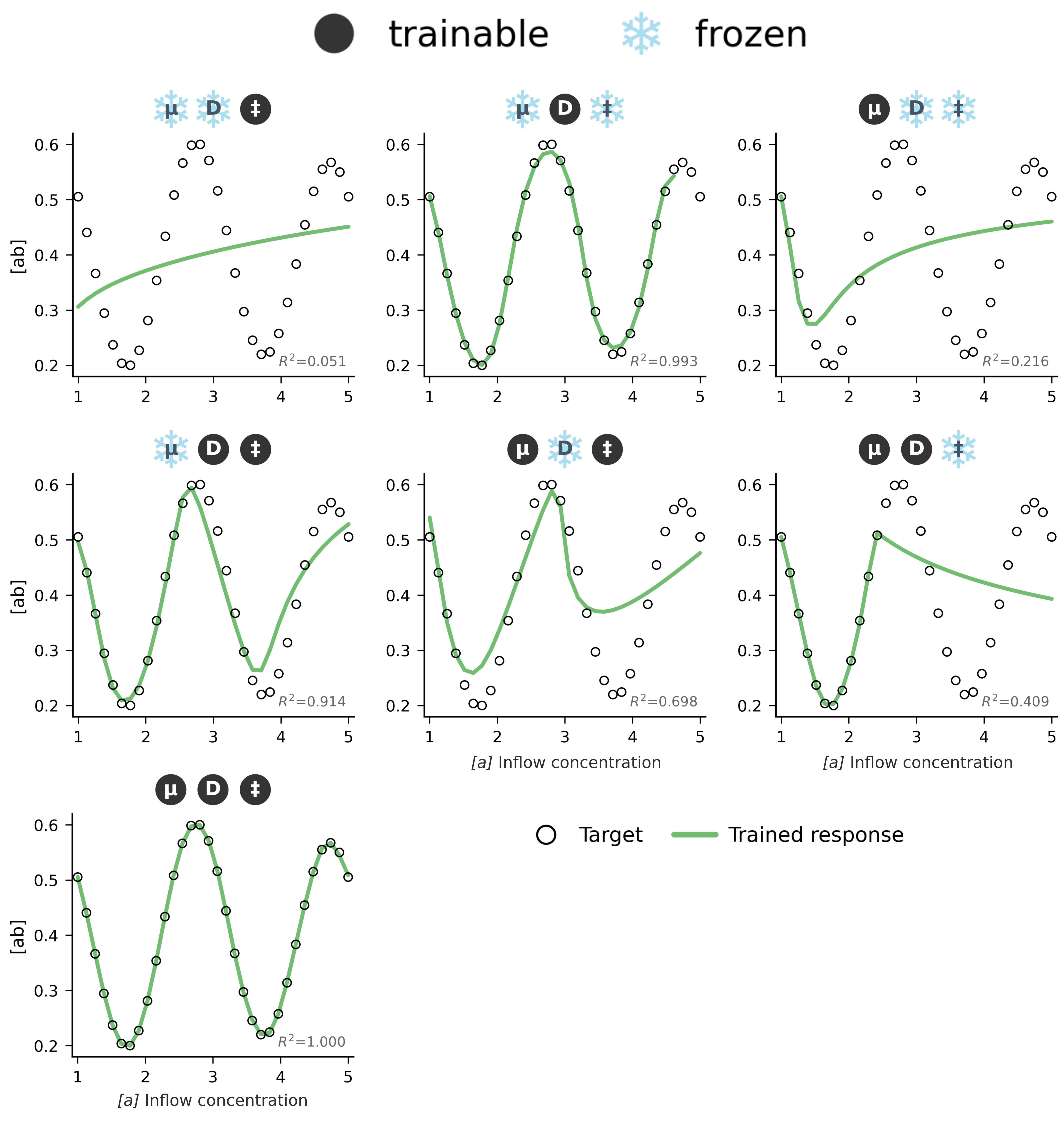}
    \caption{
    Example responses on a single target $H=2$ from the parameter-freezing bins in Fig.~\ref{fig:fig3}, chosen to highlight fit quality differences among all different parameter frozen bins.
    }
    \label{fig:fig3_plotexamples}
\end{figure*}

Figure~\ref{fig:fig3_plotexamples} shows trained responses from the different parameter-freezing bins against a common target. These curves illustrate the variation in fit quality underlying the aggregate success rates in Fig.~\ref{fig:fig3}.

The examples are provided for visualization and are not additional trials beyond the \(128\) targets used to calculate the success rates.

\section{Implementation and Reproducibility}
\label{apx:reproducibility}

\subsection{Software, Numerical Precision, and Hardware}
\label{apx:software}

The CRN dynamics, steady-state solver, implicit gradients, optimization, and ODE verification were implemented in JAX \cite{jax2018github} and Diffrax \cite{kidger2021on}. All calculations used 64-bit floating-point precision. The scaling and parameter-freezing studies were run on T4 NVIDIA GPUs and independent training runs were parallelized across targets and parameter bins. No state was shared between runs except where the same target ensemble was intentionally reused.

\subsection{Numerical Settings}
\label{apx:numericalsettings}

Table~\ref{tab:numerical_settings} collects the numerical settings used across all experiments. Values specific to one study are reported in the corresponding study section.
\begin{table*}[t]
      \caption{Numerical settings used for training and verification.}
      \label{tab:numerical_settings}
      \centering
      \begin{ruledtabular}
      \begin{tabular}{lll}
          Quantity & Value & Description \\
          \hline
          \(K\) &
          \(32\) &
          Number of input points \\
          \([x_{\min},x_{\max}]\) &
          \([1,5]\) &
          Input interval \\
          \([f_{\min},f_{\max}]\) &
          \([0.2,0.6]\) &
          Target vertical interval \\
          \(\epsilon_c\) &
          \(10^{-15}\) &
          Positive concentration floor \\
          \(\Delta t_0\) &
          \(10^{-2}\) &
          Initial pseudo-time step \\
          \(\Delta t_{\max}\) &
          \(10^{8}\) &
          Maximum pseudo-time step \\
          \(g\) &
          \(2\) &
          Pseudo-time step growth cap \\
          \(\epsilon_{\mathrm{abs}}\) &
          \(10^{-8}\) &
          Root-finder absolute tolerance \\
          \(\epsilon_{\mathrm{rel}}\) &
          \(10^{-7}\) &
          Root-finder relative tolerance \\
          \(\epsilon_{\mathrm{trust}}\) &
          \(10^{-6}\) &
          Trusted-gradient threshold \\
          \(M_{\mathrm{PTC}}\) &
          \(150\) &
          Root-finder iteration limit \\
          \(g_{\max}\) &
          \(10\) &
          Gradient clipping threshold \\
          \(\eta\) &
          \(6\times10^{-3}\) &
          Adam learning rate \\
          \((\beta_1,\beta_2)\) &
          \((0.98,0.999)\) &
          Adam moment parameters \\
          \(\lambda_{\mathrm{res}}\) &
          \(10\) &
          Residual penalty coefficient \\
          \(\lambda_{\mathrm{eig}}\) &
          \(10^{-2}\) &
          Spectral penalty coefficient \\
          \(\lambda_{\mathrm{cond}}\) &
          \(0.4\) &
          Conditioning penalty coefficient \\
          \(\chi_{\mathrm{thresh}}\) &
          \(9\) &
          Log-condition-number threshold \\
          \(E_{\max}\) &
          \(16{,}000\) &
          Maximum training epochs \\
          \(\mathrm{atol}\), Kvaerno5 &
          \(10^{-6}\) &
          ODE absolute tolerance \\
          \(\mathrm{rtol}\), Kvaerno5 &
          \(10^{-6}\) &
          ODE relative tolerance \\
          \(t_{\max}\) &
          \(5\times10^{6}\) &
          Maximum ODE integration time \\
          \(\max_i|F_i|\) &
          \(10^{-4}\) &
          ODE convergence threshold \\
      \end{tabular}
      \end{ruledtabular}
  \end{table*}

\subsection{Random Seeds and Run Indexing}
\label{apx:randomness}

Random numbers are used to generate target functions and initialize trainable parameters. Target generation and parameter initialization use separate random streams so that either can be reproduced independently.

Each run is identified by its study, topology or parameter bin, target complexity, and target index. The target seed is derived deterministically from the position in the experimental grid:
\begin{equation}
  s_{\mathrm{target}} = 500{,}000{,}042 + i_{\mathrm{row}}\times 100{,}003 + i_{\mathrm{poly}}\times 7{,}932,
\end{equation}
where \(i_{\mathrm{row}}\) indexes the target complexity level and
\(i_{\mathrm{poly}}\) indexes the polynomial within that level. The
initialization seed is derived from the linear experiment index:
\begin{equation}
  s_{\mathrm{init}} = 42 + i_{\mathrm{exp}}\times 13.
\end{equation}
The base offsets and strides are coprime, ensuring no seed collisions across
the grid. Both seeds are recorded in the released configuration files.

Target draws are shared across topologies (Fig.~\ref{fig:fig2}) and across
parameter bins (Fig.~\ref{fig:fig3}), so that each network architecture or
parameter subset is evaluated on an identical set of target functions.
Initialization seeds differ per experiment, giving each topology--target pair
an independent random starting point.

\subsection{Code Availability}
\label{apx:codeavailability}

Example code for generating the reaction networks, constructing target ensembles, training the models and performing dynamical verification is available at
\url{https://github.com/dayhofflabs/chemical-learning}

\clearpage

\end{document}